\documentclass[letterpaper, 10 pt, conference]{ieeeconf}  

\IEEEoverridecommandlockouts                              

\usepackage{cite}
\usepackage{caption}
\usepackage{amsmath,amssymb,amsfonts}
\usepackage{algorithmic}
\usepackage{graphicx}
\usepackage{subcaption}
\usepackage{multirow}
\usepackage{booktabs}
\usepackage{textcomp}
\usepackage{xcolor}
\usepackage{makecell}
\usepackage{balance}

\title{\LARGE \bf
Reliable iToF Depth Sensing via Sensor-Intrinsic Uncertainty Modeling and State-Space Restoration
}

\author{Yansong Du, Yutong Deng, Yuting Zhou, Zhancong Xu, Yingjia Lu, Mengdi Wang, Feiyu Jiao, Bangyao Wang,\\Zhaoxiang Jiang, and Xun Guan%
\thanks{Yansong Du and Yutong Deng contributed equally to this work.}%
\thanks{This work was supported in part by the Guangdong Innovative and Entrepreneurial Research Team under Grant 2023ZT10X137, the National Natural Science Foundation of China under Grant U23A20282, and the Shenzhen Science and Technology Program under Grants KJZD20240903095600001 and ZDCY20250901110702003. Corresponding authors: Zhaoxiang Jiang and Xun Guan.}%
\thanks{Yansong Du, Yutong Deng, Yuting Zhou, Zhancong Xu, Yingjia Lu, Mengdi Wang, Feiyu Jiao, Bangyao Wang, and Xun Guan are with the Tsinghua Shenzhen International Graduate School, Tsinghua University, Shenzhen 518055, China.}%
\thanks{Zhaoxiang Jiang is with the Guangdong Laboratory of Artificial Intelligence and Digital Economy (SZ), Shenzhen 518060, China.}%
}

\begin{document}

\maketitle
\thispagestyle{empty}
\pagestyle{empty}

\begin{abstract}
Indirect time-of-flight (iToF) cameras provide compact and cost-effective dense depth measurements, but their ranging accuracy is often degraded by sensor-intrinsic uncertainty under practical imaging conditions. Spatially uniform or range-only Gaussian perturbations cannot accurately reproduce the range-dependent and signal-dependent noise characteristics of real iToF measurements, leading to a synthetic-to-real gap for learning-based restoration. To address this problem, we propose a joint depth-uncertainty modeling and restoration framework for reliable iToF sensing. A sensor-intrinsic depth-uncertainty model is first developed from calibrated tap responses, returned-signal levels, and sensor noise statistics through a depth-oriented weighted least-squares formulation. The resulting pixel-wise uncertainty is used for heteroscedastic depth synthesis and uncertainty-aware restoration supervision. Based on this heteroscedastic data synthesis, we further develop a U-shaped restoration network with Depth Visual State Space (DVSS) blocks, which combine long-range state-space modeling with convolutional spatial-channel refinement for structure-preserving depth recovery. Experiments on synthetic data and measurements captured by an in-house iToF prototype validate the proposed uncertainty model under varying range and returned-signal conditions. Controlled comparisons with fixed and range-aware Gaussian noise, together with evaluations on U-Net, Restormer, and DVSS, further demonstrate that the proposed synthesis consistently benefits different restoration backbones. The complete framework achieves 40.85~dB PSNR and 2.54~mm MAE on the synthetic test set, and 35.42~dB PSNR and 4.87~mm MAE on real iToF measurements.
\end{abstract}

\noindent\textbf{Keywords:} 
Indirect time-of-flight, depth restoration, uncertainty modeling, heteroscedastic noise, state-space model.

\section{INTRODUCTION}

Indirect time-of-flight (iToF) cameras are widely used for robotic perception, machine vision, and human machine interaction, owing to their compact form factor, low cost, and compatibility with mature CMOS imaging technology~\cite{qiao2024rgb,du2026structure}. By estimating scene depth from the phase delay between emitted and received modulated optical signals, iToF sensors provide dense metric measurements for geometric reconstruction, autonomous localization, mapping, and closed-loop decision-making~\cite{kazerouni2022survey,sousa2023systematic}. However, their reliability is still limited by sensor-intrinsic uncertainty under practical sensing conditions. Random fluctuations in tap-level correlation counts, dark and read noise, non-ideal modulation responses, and residual calibration errors propagate to phase estimation and eventually appear as depth-domain uncertainty. In this work, we focus on the stochastic component of this uncertainty, especially the range-dependent and signal-dependent depth variance under weak returned signals, low-reflectance surfaces, or long-range measurements.

Existing studies have investigated iToF error suppression through physical modeling and calibration~\cite{du2025modeling,du2025random,du2025calibration}, as well as sensor correction and complementary reconstruction~\cite{du2025new,deng2025monocular}. Analytical methods provide insight into tap-level or phase-domain noise propagation, but they are often simplified for depth-domain data generation, for example by assuming spatially uniform or range-only Gaussian perturbations. Such assumptions cannot accurately describe real iToF measurements, where depth uncertainty depends on photon-counting statistics, sensor noise floors, range, and local returned signal strength. Learning-based restoration methods can improve depth quality by learning nonlinear correction mappings, but their performance depends heavily on whether the training data matches real sensor statistics. This mismatch leads to a synthetic-to-real distribution gap and limits robustness across devices, distances, and acquisition conditions.

To improve the reliability of iToF depth sensing, this paper studies depth restoration from a sensor-uncertainty perspective. Instead of treating depth noise as a fixed image-level perturbation, we derive a per-pixel depth uncertainty model from tap-level iToF statistics. The tap measurements are modeled by considering both signal-dependent counting noise and signal-independent sensor noise, and a depth-oriented weighted least-squares formulation is used to propagate measurement uncertainty to the final depth estimate. The resulting uncertainty map enables heteroscedastic noisy depth synthesis with spatially varying random errors consistent with real iToF measurements, and is further used for noise-aware restoration training.

\begin{figure}[!t]
  \centering
  \includegraphics[
    width=0.96\columnwidth,
    height=0.24\textheight,
    keepaspectratio
  ]{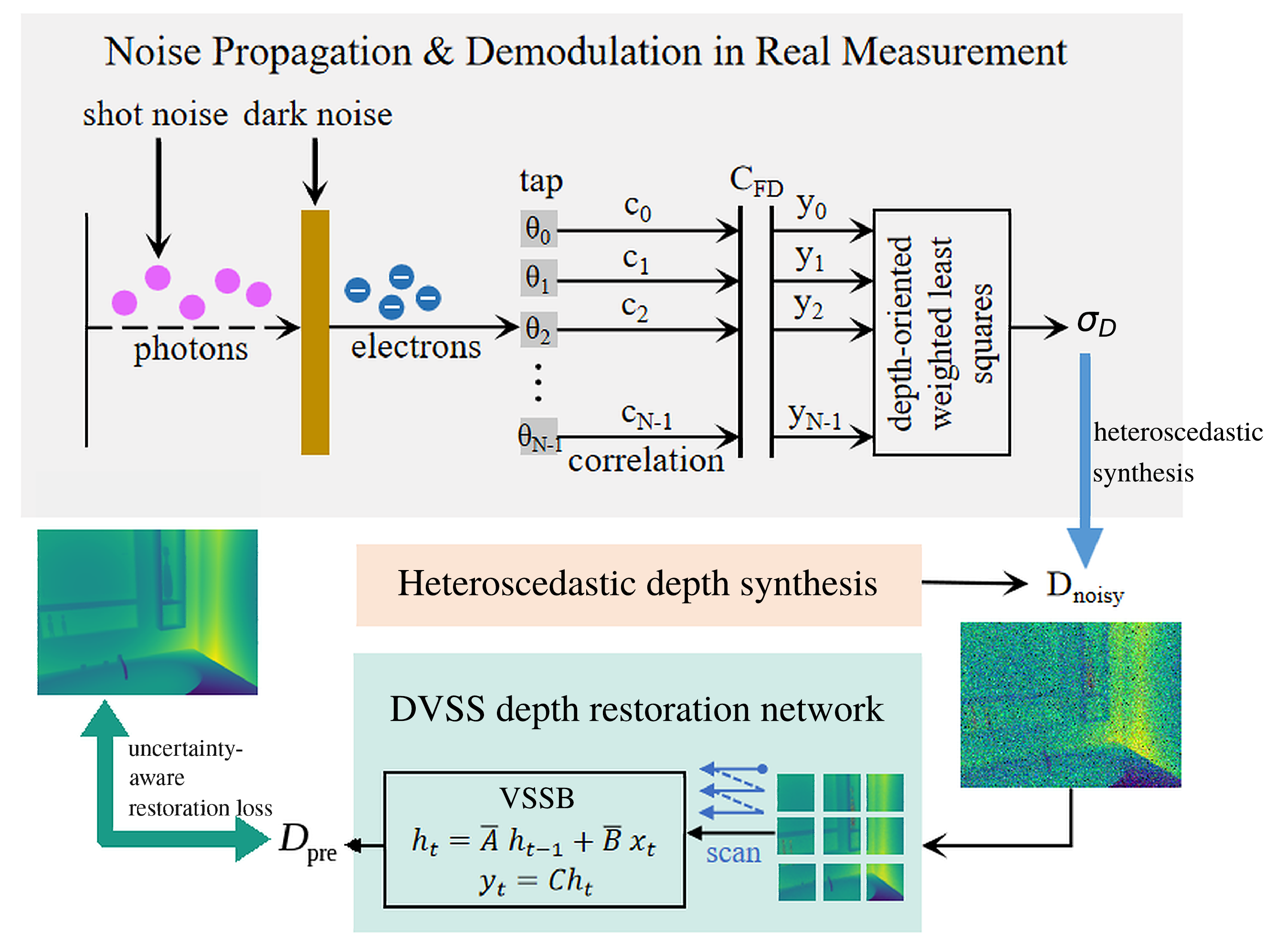}
  \caption{Overall workflow of the proposed reliable iToF depth sensing framework. Sensor-intrinsic uncertainty modeling produces heteroscedastic noisy-depth observations and uncertainty-aware supervision, which are used to train the restoration network equipped with DVSS blocks for accurate and structure-preserving depth recovery.}
  \label{fig:overall_workflow}
  \vspace{-8pt}
\end{figure}

Based on the proposed sensor-intrinsic depth-uncertainty model, we design a DVSS-based depth restoration framework for iToF measurements. The estimated uncertainty is used for heteroscedastic data synthesis and uncertainty-aware training supervision, encouraging the network to suppress unreliable depth regions while preserving reliable geometric structures. The network adopts a U-shaped architecture with a Depth Visual State Space (DVSS) block, where the state-space branch captures long-range structural dependencies and the convolutional spatial-channel refinement branch preserves local geometric details and depth discontinuities. This design couples sensor-level uncertainty modeling with structure-preserving depth restoration~\cite{song2022all,tang2022ghostnetv2}. Its long-range branch further benefits from efficient state-space modeling~\cite{gu2024mamba,wang2025mamba}. The overall workflow of the proposed framework is summarized in Fig.~\ref{fig:overall_workflow}.

The main contributions of this work are summarized as follows:

\begin{itemize}
    \item We propose a sensor-intrinsic depth-uncertainty model for iToF sensors. The model derives per-pixel depth uncertainty from signal-dependent counting statistics and sensor-intrinsic noise through a depth-oriented weighted least-squares formulation, enabling heteroscedastic noisy depth synthesis that reflects the range-dependent and signal-dependent stochastic behavior of real iToF measurements.
    
    \item We develop a DVSS-based iToF depth restoration network coupled with the proposed heteroscedastic synthesis. The propagated uncertainty is used to construct heteroscedastic noisy depth observations and uncertainty-aware training weights, while the proposed DVSS block combines state-space global modeling with convolutional spatial-channel refinement for accurate and structure-preserving depth recovery.
    
    \item We conduct experiments on synthetic data and real measurements captured by an in-house iToF prototype. Range- and returned-signal-dependent measurements validate the proposed uncertainty model, while controlled comparisons across U-Net, Restormer, and DVSS demonstrate that the benefit of the proposed synthesis is independent of a specific restoration architecture.
    
\end{itemize}

\section{Related Work}

\textbf{iToF error modeling and sensor-level correction.} Indirect time-of-flight (iToF) cameras recover depth from phase-shifted correlation measurements, and their accuracy is affected by sensor noise and scene-dependent artifacts. Prior studies have analyzed lock-in ToF camera principles, calibration issues, shot-noise-induced distance uncertainty, and background illumination effects~\cite{foix2011lockin,illade2015shotnoise}. To improve modeling fidelity, ToF simulators have been developed at the illumination and sensor levels, including AMCW simulation with quantified multipath effects and sensor charge modeling~\cite{lambers2015simulation,bulczak2018quantified}. Hardware- and acquisition-side correction methods have also been explored for depth-frame difference sensing, motion-blur compensation, and coding-function design~\cite{gao2023deblurring,gutierrez2019practical,li2022fisher}. Our previous work further characterized the propagation of signal-light, ambient-light, and dark-noise fluctuations through phase demodulation and system correction, and used the resulting phase-domain model for waveform-duty-cycle optimization~\cite{du2025modeling}. In contrast, the present work develops a restoration-oriented depth uncertainty formulation based on a local WLS information matrix and investigates its use for pixel-wise heteroscedastic synthesis and learning-based depth restoration. Existing physical modeling and correction methods improve specific noise or artifact sources, but their effectiveness is often tied to camera configuration, calibration, or acquisition assumptions, making residual noise difficult to model under diverse materials, illumination conditions, and scene geometries.

\textbf{Learning-based ToF denoising and the synthetic-to-real gap.} Deep learning has become an effective tool for ToF reconstruction because it can combine denoising, MPI suppression, and phase/depth refinement in a data-driven manner. DeepToF and subsequent end-to-end pipelines demonstrated that CNNs can recover depth from raw or intermediate ToF measurements~\cite{marco2017deeptof,su2018deeptof}. Other works introduced synthetic ToF datasets, task-specific networks, attention mechanisms, and multi-frequency or transient representations to improve robustness~\cite{guo2018flat,wang2022attentiongan,gutierrez2021itof2dtof}. More recent studies exploit domain adaptation and real noisy measurements to reduce dependence on ideal clean supervision~\cite{agresti2019uda,schelling2022radu,zhang2024dcs2noise}. Multi-frame methods further leverage temporal correlations for stable depth recovery~\cite{dong2024dualcorrelation,wang2025giga}. Statistical approaches have also incorporated sensor-noise distributions into iToF restoration. BayesToF derives a multiresolution Bayesian MMSE estimator from a multinomial-Poisson likelihood and a Gaussian-scale-mixture prior, directly suppressing photon and thermal noise in raw iToF measurements without network training~\cite{idoughi2026bayestof}. In contrast, our method converts calibrated tap statistics into pixel-wise depth uncertainty for heteroscedastic training-data synthesis and evaluates its transferability across different restoration backbones. Nevertheless, real iToF noise is governed by coupled factors including photon statistics, demodulation non-idealities, material reflectance, motion, and scene geometry. Spatially uniform or range-only synthetic noise injection often cannot reproduce these coupled effects, leading to a synthetic-to-real distribution shift and unstable deployment performance.

\textbf{Efficient long-range modeling for depth denoising.} The network backbone is also critical for practical iToF denoising. Depth errors produced by iToF cameras are not purely local: MPI, motion blur, flying pixels, and low-SNR regions often exhibit spatially correlated structures and global consistency constraints. CNNs are efficient but may require deeper or wider networks to capture long-range context~\cite{tang2022ghostnetv2}. Transformer-style attention can model global dependencies, yet its memory and computational costs become expensive for high-resolution dense prediction~\cite{cai2023efficientvit}. Recently, selective state-space models such as Mamba have provided a linear-time alternative for long-sequence modeling~\cite{gu2024mamba}, and Mamba-style modeling has also been explored for iToF MPI correction~\cite{an2025mpi}. Motivated by these observations, our work focuses on physically faithful noise modeling and efficient long-range feature aggregation for robust iToF depth denoising under practical deployment constraints.

\section{Method}

The proposed method couples sensor-intrinsic depth-uncertainty modeling with a DVSS-based depth restoration framework. Instead of injecting scene-independent Gaussian perturbations into clean depth maps, the proposed synthesis pipeline derives pixel-wise depth uncertainty from tap-level measurement statistics. Specifically, the clean depth $D$, signal level $a$, and sensor parameters are used to construct the tap-level response, measurement mean, and covariance. A depth-oriented weighted least-squares formulation is then linearized to estimate the local depth variance, which determines the pixel-wise perturbation scale for heteroscedastic noisy depth synthesis. The synthesized noisy depth maps are used to train and evaluate a U-shaped denoising network equipped with the proposed Depth Visual State Space (DVSS) blocks for structure-preserving depth restoration.

\subsection{iToF Tap Response Model}

For each valid pixel location $p$, let $D(p)$ denote the clean depth and $a(p)$ denote the scaled returned-signal level used by the tap-level noise model. In real measurements, this signal level corresponds to the detected photoelectron level after sensor conversion, while in synthetic data it is approximated from the ToF amplitude and rescaled to the signal-level range used for noise synthesis. The signal level $a(p)$ is affected by the incident illumination, surface reflectance, optical attenuation, exposure setting, and sensor conversion efficiency. To keep the notation concise, the pixel index $p$ is omitted in the following derivation unless otherwise specified.

Within one iToF modulation cycle, the sensor acquires $N$ phase-shifted tap measurements. The phase shift of the $n$-th tap is defined as
\begin{equation}
\theta_n = \frac{2\pi n}{N}, \quad n=0,\ldots,N-1 .
\end{equation}
For a single modulation frequency, the depth-dependent phase delay is given by
\begin{equation}
\psi(D) = \frac{4\pi f}{c_{\mathrm{light}}}D ,
\end{equation}
where $f$ is the modulation frequency and $c_{\mathrm{light}}$ is the speed of light.

The systematic phase nonlinearity and fixed-pattern bias are assumed to be compensated by the device-level calibration pipeline. Therefore, this work focuses on the stochastic random error after calibration. Let $c(\cdot)$ denote the calibrated tap response function. The response of the $n$-th tap to depth $D$ is modeled as
\begin{equation}
r_n(D) = c\!\left(\psi(D)-\theta_n\right).
\end{equation}
Using the signal level $a$, the expected electron count of the $n$-th tap is written as
\begin{equation}
\mu_n(D,a) = a \cdot r_n(D) + \mu_{\mathrm{dark}},
\end{equation}
where $\mu_{\mathrm{dark}}$ denotes the calibrated mean dark offset, which is treated as a deterministic component of the tap expectation. The corresponding tap measurement is
\begin{equation}
y_n = \mu_n(D,a) + \eta_n ,
\end{equation}
where $\eta_n$ denotes the random measurement fluctuation.

Under photon-counting noise with a signal-independent sensor noise floor, the variance of the $n$-th tap measurement is modeled as
\begin{equation}
\sigma_n^2 = a \cdot r_n(D) + \sigma_{\mathrm{dark}}^2 .
\end{equation}
The first term models signal-dependent shot noise, whereas the calibrated offset $\mu_{\mathrm{dark}}$ is treated as deterministic and $\sigma_{\mathrm{dark}}^2$ accounts for the residual signal-independent dark/readout noise variance around this offset. This formulation explicitly models the heteroscedastic nature of iToF measurements: the tap-level uncertainty changes with both the local signal level and the depth-dependent response. All sensor-specific parameters are determined from an independent calibration sequence before noise synthesis. The calibrated tap response $c(\cdot)$ is obtained from a phase sweep of a uniform planar target by averaging repeated tap measurements at each phase position. The mean dark offset $\mu_{\mathrm{dark}}$ and the residual dark/read-noise variance $\sigma_{\mathrm{dark}}^{2}$ are estimated from 100 emitter-off captures under the same sensor configuration. For synthetic data, the valid ToF amplitude is first normalized and then mapped to the empirical returned-signal interval measured from the calibration sequence, i.e., $a(p)=a_{\min}+(a_{\max}-a_{\min})\widetilde{A}(p)$, where $a_{\min}$ and $a_{\max}$ are determined from the lower and upper percentiles of valid returned-signal measurements. All calibration parameters are fixed before training and are not estimated from the validation or test scenes.

\subsection{Depth-Oriented Weighted Least-Squares Noise Modeling}

The $N$ tap measurements at a pixel are collected into a vector
\begin{equation}
y = [y_0,y_1,\ldots,y_{N-1}]^{T}.
\end{equation}
The corresponding tap expectation vector is
\begin{equation}
\mu(D,a) =
[\mu_0(D,a),\mu_1(D,a),\ldots,\mu_{N-1}(D,a)]^{T}.
\end{equation}
According to the tap variances, the measurement covariance matrix is defined as
\begin{equation}
\Sigma_y =
\operatorname{diag}
\left(
\sigma_0^2,\sigma_1^2,\ldots,\sigma_{N-1}^2
\right).
\end{equation}

Instead of explicitly estimating phase as an intermediate variable, the proposed formulation directly treats depth as the target variable. To characterize the local uncertainty of depth and returned-signal strength, we consider the following depth-oriented weighted least-squares objective:
\begin{equation}
(\hat{D},\hat{a})
=
\arg\min_{D,a>0}
\left(
y-\mu(D,a)
\right)^{T}
\Sigma_y^{-1}
\left(
y-\mu(D,a)
\right).
\end{equation}
Here, $a$ is included as an auxiliary signal-level variable. During noise synthesis, the clean depth map and the signal-level map define the operating point at which the local uncertainty is evaluated.

Let
\begin{equation}
\xi = [D,a]^{T}.
\end{equation}
The Jacobian of the measurement model is
\begin{equation}
J_{\xi}
=
\frac{\partial \mu}{\partial \xi}
=
\left[
\frac{\partial \mu}{\partial D},
\frac{\partial \mu}{\partial a}
\right].
\end{equation}
For the $n$-th tap, the two partial derivatives are
\begin{equation}
\frac{\partial \mu_n}{\partial D}
=
a\,
c'\!\left(\psi(D)-\theta_n\right)
\frac{4\pi f}{c_{\mathrm{light}}},
\end{equation}
and
\begin{equation}
\frac{\partial \mu_n}{\partial a}
=
r_n(D).
\end{equation}

By locally linearizing the weighted least-squares estimator around the operating point, the covariance of the estimated parameter vector can be approximated as
\begin{equation}
\operatorname{Cov}(\hat{\xi})
\approx
\left(
J_{\xi}^{T}
\Sigma_y^{-1}
J_{\xi}
\right)^{-1}.
\end{equation}
The pixel-wise depth variance is the component of this covariance matrix corresponding to the depth variable:
\begin{equation}
\sigma_D^2
=
e_D^{T}
\left(
J_{\xi}^{T}
\Sigma_y^{-1}
J_{\xi}
\right)^{-1}
e_D,
\quad
e_D=[1,0]^{T}.
\end{equation}
The WLS formulation is used to evaluate local depth uncertainty around the calibrated operating point and does not replace the camera's standard phase-demodulation and depth-decoding pipeline. This expression shows that the depth uncertainty is jointly governed by the local signal level, the depth sensitivity of the tap response, and the sensor noise variance. A larger response sensitivity reduces the depth variance, whereas weak signal levels or a higher noise floor increase the final uncertainty.

\subsection{Physics-Guided Noisy Depth Synthesis}

The depth variance derived above is evaluated for each valid pixel. Restoring the pixel index $p$, the local depth standard deviation is
\begin{equation}
\sigma_D(p)=\sqrt{\sigma_D^2(p)}.
\end{equation}
The noisy depth observation is then synthesized as
\begin{equation}
D_{\mathrm{noisy}}(p)
=
D(p)+\sigma_D(p)\epsilon(p),
\quad
\epsilon(p)\sim\mathcal{N}(0,1).
\end{equation}
This formulation does not assign a fixed global noise level to the entire depth map. Instead, each pixel has its own uncertainty scale determined by the iToF tap response, signal-level map, and sensor noise statistics. Although $\epsilon(p)$ is sampled independently across pixels, the standard deviation map $\sigma_D(p)$ is spatially structured because it depends on scene depth and signal strength. Therefore, the synthesized noisy depth reflects the common iToF behavior that low-signal, low-reflectance, or high-uncertainty regions exhibit stronger random depth errors. This work focuses on the stochastic depth uncertainty after device-level phase calibration, and does not explicitly synthesize a separate depth-domain bias map.

\subsection{Denoising Network Architecture}

\begin{figure*}[!t]
  \centering
  \includegraphics[width=0.95\textwidth]{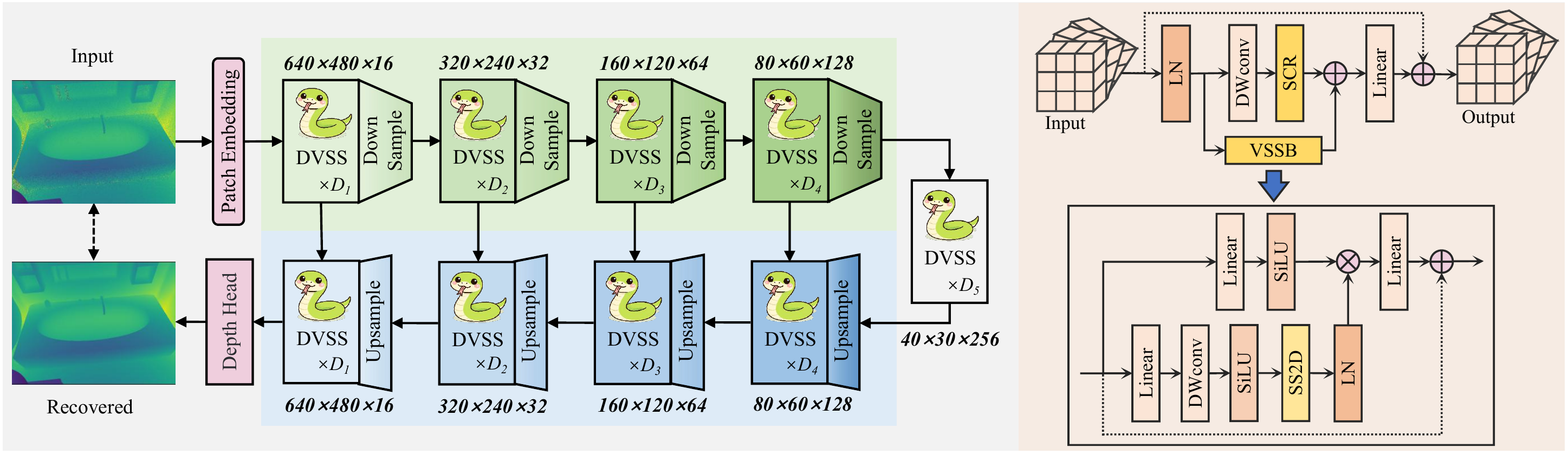}
  \caption{Overview of the proposed denoising network and its core modules. The left side shows the U-shaped architecture with stacked DVSS blocks and skip connections. The right side illustrates the proposed DVSS block and its VSSB submodule.}
  \label{fig:network}
\end{figure*}

Given the synthesized noisy depth map $D_{\mathrm{noisy}}$, the proposed network predicts the restored depth map $\hat{D}$. As shown in Fig.~\ref{fig:network}, the denoising backbone follows a U-shaped architecture equipped with stacked DVSS blocks. The input depth map is first projected into the feature space by a Patch Embedding layer. The encoder progressively extracts multi-scale features through downsampling stages and stacked DVSS blocks, while the bottleneck aggregates high-level contextual representations. The decoder restores the spatial resolution through symmetric upsampling stages and uses skip connections to recover fine geometric details from encoder features. A Depth Head finally maps the restored features to the denoised depth output.

The proposed Depth Visual State Space (DVSS) block is the core component of the network. After layer normalization, the input feature $X$ is processed by two parallel branches. The global branch employs a Visual State Space Block (VSSB) to model long-range dependencies in 2D depth features:
\begin{equation}
X_g =
\operatorname{VSSB}
\left(
\operatorname{LN}(X)
\right).
\end{equation}
The local branch applies depthwise convolution followed by a Spatial-Channel Refinement (SCR) module to enhance local structures and channel responses:
\begin{equation}
X_l =
\operatorname{SCR}
\left(
\operatorname{DWConv}
\left(
\operatorname{LN}(X)
\right)
\right).
\end{equation}
The two branches are fused through channel concatenation and a $1\times1$ convolution with a residual connection:
\begin{equation}
Y =
X+
\operatorname{Conv}_{1\times1}
\left(
[X_g,X_l]
\right),
\end{equation}
where $[\cdot,\cdot]$ denotes channel-wise concatenation. The VSSB branch focuses on efficient global modeling, while the convolutional SCR branch refines local spatial structures and channel responses.

\begin{figure}[!t]
  \centering
  \includegraphics[width=0.72\linewidth]{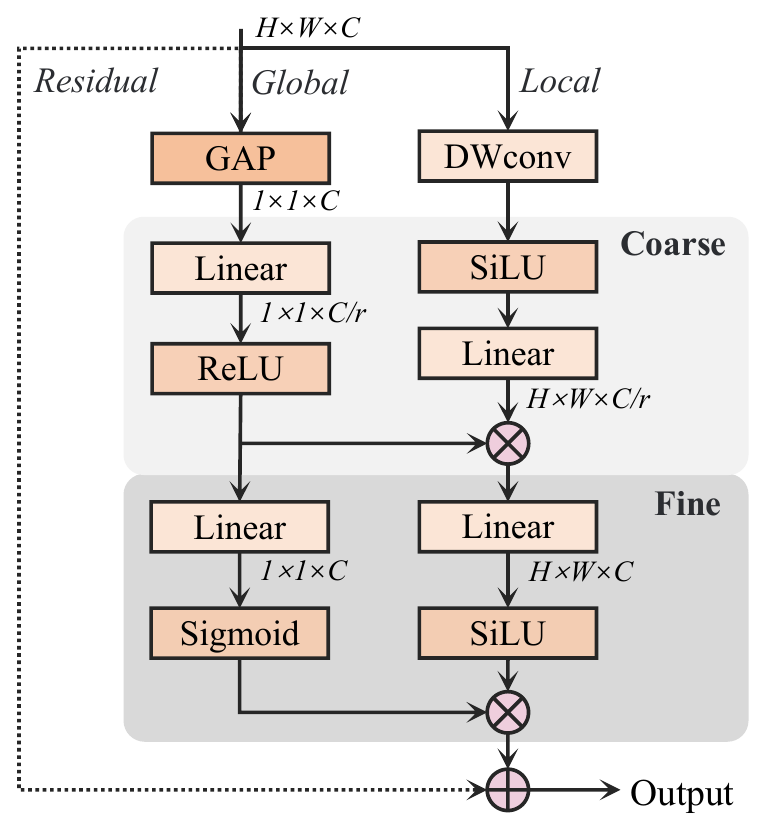}
  \caption{Illustration of the proposed SCR module. A global branch and a local branch interact in a coarse-to-fine manner to jointly refine channel and spatial responses.}
  \label{fig:scr}
\end{figure}

As illustrated in Fig.~\ref{fig:scr}, the SCR module introduces a global-local refinement mechanism to improve feature selectivity. The global branch provides channel-level context, while the local branch preserves spatially detailed responses. Their interaction allows the network to suppress stochastic depth noise while maintaining local edges, thin structures, and depth discontinuities. The combination of state-space global modeling and SCR-based local refinement enables robust and structure-preserving iToF depth denoising.

\section{Noise Model Validation}

\begin{figure}[!htbp]
  \centering
  \captionsetup{font=footnotesize} 
  \includegraphics[width=\linewidth]{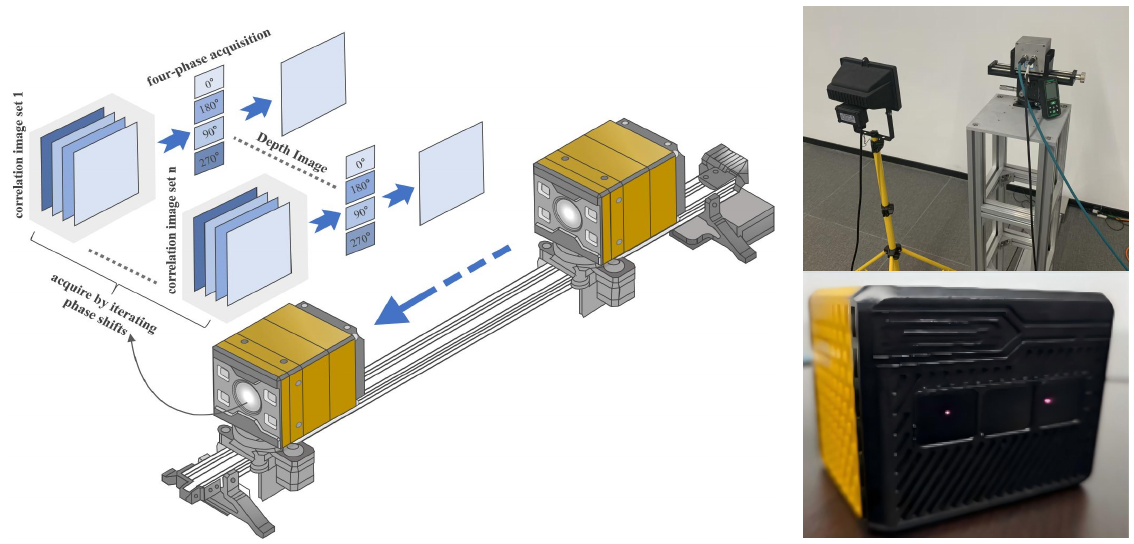}
  \caption{Experimental setup and in-house iToF imaging system, including the acquisition configuration, camera module, and measurement rig.}
  \label{fig:fig5}
  \vspace{-10pt}
\end{figure}

Our proposed noise model is validated through real-scene acquisition experiments using an in-house developed iToF camera and the associated measurement system, as illustrated in Fig.~\ref{fig:fig5}. 

A planar target is measured over a working range from 500\,mm to 1500\,mm, with validation performed at every 100\,mm interval. At each distance, we estimate the empirical temporal depth-noise standard deviation from repeated real iToF captures and average it over the valid region of interest (ROI). The real statistics are compared with a range-aware Gaussian baseline and the proposed sensor-intrinsic depth-uncertainty model. The range-aware Gaussian baseline uses a distance-dependent noise standard deviation, but remains spatially uniform within each distance. Specifically, its standard deviation is modeled by a second-order function $\sigma_{\mathrm{range}}(D)=\alpha_0+\alpha_1D+\alpha_2D^2$, whose coefficients are estimated by least-squares fitting on an independent planar calibration sequence. The calibration captures used to determine $\alpha_0$, $\alpha_1$, and $\alpha_2$ are separated from the measurements used for the range- and returned-signal-dependent validation experiments. In contrast, the proposed model directly evaluates pixel-wise depth uncertainty from calibrated tap statistics through the depth-oriented WLS formulation, allowing the synthesized noise to adapt to both range and local returned-signal strength.

\begin{figure*}[t]
  \centering
  \includegraphics[width=0.325\textwidth]{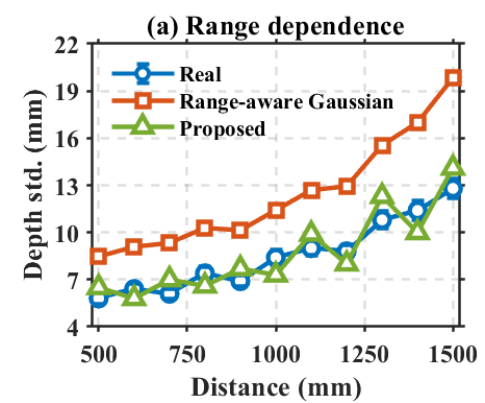}\hfill
  \includegraphics[width=0.325\textwidth]{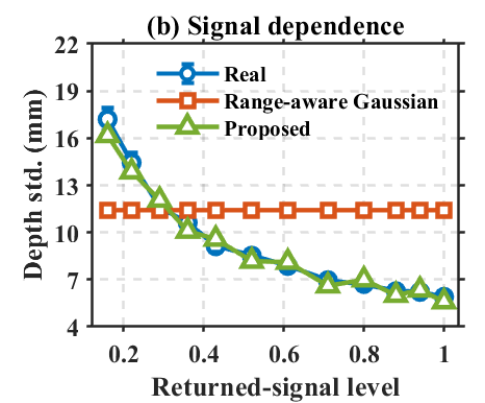}\hfill
  \includegraphics[width=0.325\textwidth]{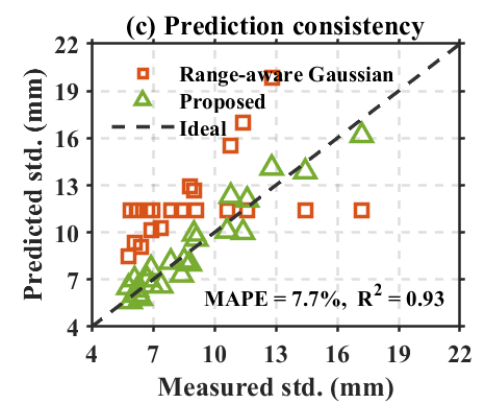}
  \caption{Validation of real-camera depth uncertainty. (a) Temporal depth standard deviation over the 500-1500~mm range. (b) Depth uncertainty under varying returned-signal levels at a fixed distance. (c) Predicted versus measured depth standard deviation over all evaluated conditions.}
  \label{fig:fig6}
  \vspace{-3mm}
\end{figure*}

Fig.~\ref{fig:fig6} reports the quantitative validation results over the full evaluated range. Fig.~\ref{fig:fig6}(a) shows the mean depth-noise standard deviation as a function of distance. The real temporal noise increases with range, which is consistent with the reduced return signal at longer distances. The proposed model closely follows this increasing trend and remains near the real measurements across the tested range. By comparison, although the range-aware Gaussian baseline also increases with distance, it systematically deviates from the real temporal noise because its spatially uniform assumption cannot capture the signal-dependent uncertainty of iToF measurements.

To isolate signal dependence from range, we further fix the target distance at 1500~mm and vary the returned-signal level using coplanar black, gray, and white surfaces under four acquisition settings. For each condition, the empirical temporal depth standard deviation is computed from 100 repeated captures. Because the range-aware Gaussian baseline assigns a common uncertainty to pixels at the same distance, it cannot reproduce the measured variation across return levels. In contrast, the proposed model closely follows the reduction in depth uncertainty as the returned signal increases.

We further pool all range and returned-signal conditions to compare the predicted and measured depth standard deviations. The proposed model achieves a MAPE of 7.7\% and an $R^2$ of 0.93, substantially outperforming the range-aware Gaussian baseline. These results confirm that both distance and local returned-signal strength are required to characterize real iToF depth uncertainty.

\section{Evaluation of the Proposed Network}

\subsection{Training Details}

We use ToF-FlyingThings3D\cite{qiu2019deep} as the primary dataset for training and evaluation. The dataset contains 6,250 synthetic samples rendered from different viewpoints, and each sample provides aligned RGB images, ToF amplitude, ToF depth, and ground-truth depth at a resolution of 640$\times$480. To construct noisy-clean depth pairs, we use the provided ToF amplitude map as the returned-signal cue for the proposed sensor-intrinsic depth-uncertainty model, since it more directly reflects the active-light response of iToF measurements than RGB luminance. The amplitude map is normalized over valid pixels and rescaled to the signal-level range used for noise synthesis, yielding the per-pixel signal-level map $a(p)$. All depth values are converted to millimeters before training and evaluation. Accordingly, MAE and MSE are reported in millimeters and squared millimeters, respectively, while PSNR is reported in decibels. PSNR is computed as $\mathrm{PSNR}=10\log_{10}(D_{\max}^{2}/\mathrm{MSE})$, where $D_{\max}$ denotes the fixed peak depth of $1000~\mathrm{mm}$ used for all test samples. The RGB image is used only for visualization and qualitative comparison, while the denoising network takes the synthesized noisy iToF depth as input and predicts the clean depth.

We randomly split the dataset into training, validation, and test sets with a ratio of 6:2:2, resulting in 3,750 training samples, 1,250 validation samples, and 1,250 test samples. All models are trained using AdamW with an initial learning rate of $2\times10^{-4}$, a batch size of 4, and 300 epochs. During training, $320\times320$ patches are randomly cropped from the original $640\times480$ depth maps, while validation and testing are performed at the full resolution. The learning rate is gradually reduced to $1\times10^{-6}$ using a cosine annealing schedule, and the checkpoint with the lowest validation MAE is used for evaluation. Unless otherwise specified, the final model is optimized with an uncertainty-aware restoration loss that combines uncertainty-weighted robust reconstruction and gradient consistency:
\begin{equation}
\mathcal{L}_{\mathrm{full}}
=
\mathcal{L}_{\mathrm{nar}}
+
\lambda_g \mathcal{L}_{\mathrm{grad}},
\end{equation}
where $\mathcal{L}_{\mathrm{nar}}$ denotes the noise-aware reconstruction term and $\mathcal{L}_{\mathrm{grad}}$ encourages local geometric consistency. Specifically, we define
\begin{equation}
\mathcal{L}_{\mathrm{nar}}
=
\frac{1}{|\Omega|}
\sum_{t \in \Omega}
w(t)\,
\rho\left(\hat{D}(t)-D_{\mathrm{gt}}(t)\right),
\end{equation}
where $\Omega$ is the set of valid pixels, $\hat{D}$ is the recovered depth, and $D_{\mathrm{gt}}$ is the ground-truth depth. The function
\begin{equation}
\rho(x)=\sqrt{x^2+\varepsilon_c^2}
\end{equation}
is the Charbonnier penalty, which is a smooth and robust approximation to the $\mathcal{L}_{1}$ loss. The uncertainty-aware weight $w(t)$ is derived from the propagated depth uncertainty of the proposed depth-uncertainty model:
\begin{equation}
w(t)
=
1+\beta \cdot \widetilde{\sigma}_{D}(t),
\end{equation}
where $\widetilde{\sigma}_{D}(t)$ denotes the normalized pixel-wise depth standard deviation. In practice, it is computed as
\begin{equation}
\widetilde{\sigma}_{D}(t)
=
\frac{\sigma_D(t)-\min_{t\in\Omega}\sigma_D(t)}
{\max_{t\in\Omega}\sigma_D(t)-\min_{t\in\Omega}\sigma_D(t)+\eta},
\end{equation}
where $\eta$ is a small constant for numerical stability. This weighting strategy encourages the network to focus more on low-signal and high-uncertainty regions, which are more vulnerable to iToF measurement uncertainty.

To further preserve structural discontinuities, we introduce a gradient consistency term:
{\small
\begin{equation}
\begin{aligned}
\mathcal{L}_{\mathrm{grad}}
&=
\frac{1}{|\Omega|}
\sum_{t \in \Omega}
\Big[
\rho\big(\nabla_x \hat{D}(t)-\nabla_x D_{\mathrm{gt}}(t)\big) \\
&\quad+
\rho\big(\nabla_y \hat{D}(t)-\nabla_y D_{\mathrm{gt}}(t)\big)
\Big],
\end{aligned}
\end{equation}
}
where $\nabla_x$ and $\nabla_y$ denote finite-difference operators along the horizontal and vertical directions, respectively. In our experiments, we set $\lambda_g=0.1$, $\beta=0.5$, $\varepsilon_c=10^{-3}$, and $\eta=10^{-6}$. For ablation analysis, we additionally report the results of the same network trained with the standard $\mathcal{L}_{1}$ reconstruction loss. Unless otherwise stated, all synthetic quantitative results are computed on the held-out test set. All experiments are conducted on a single NVIDIA RTX 4090 GPU. For controlled noise-model comparisons, the fixed Gaussian and range-aware Gaussian baselines are globally rescaled to match the average perturbation variance of the proposed synthesis over the training set, while the training samples, restoration objective, optimizer, and number of epochs are kept unchanged.

\subsection{Quantitative Comparison}

We compare the proposed method with representative iToF depth denoising approaches, including DeepToF~\cite{marco2017deeptof}, ToF-KPN~\cite{qiu2019deep}, and RCF~\cite{dong2024recurrent}. As summarized in Table~\ref{tab:quantitative_comparison}, the proposed architecture trained with the standard $\mathcal{L}_{1}$ loss already outperforms existing iToF denoising baselines, achieving 40.37~dB PSNR, 0.9985 SSIM, 91.75~mm$^2$ MSE, and 2.78~mm MAE. When trained with the proposed noise-aware restoration loss $\mathcal{L}_{\mathrm{full}}$, the final model further improves the performance to 40.85~dB PSNR, 0.9987 SSIM, 82.30~mm$^2$ MSE, and 2.54~mm MAE. Compared with RCF, the strongest baseline under the unified PSNR definition, the final model improves PSNR by 1.32~dB, improves SSIM by 0.0015, and reduces MSE and MAE by 26.1\% and 45.1\%, respectively. These improvements demonstrate that the proposed network not only enhances structural similarity but also more effectively suppresses pixel-wise ranging errors, especially in low-signal and high-uncertainty regions.

\begin{figure}[!htbp]
  \centering
  \captionsetup{font=footnotesize} 
  \includegraphics[width=0.48\textwidth]{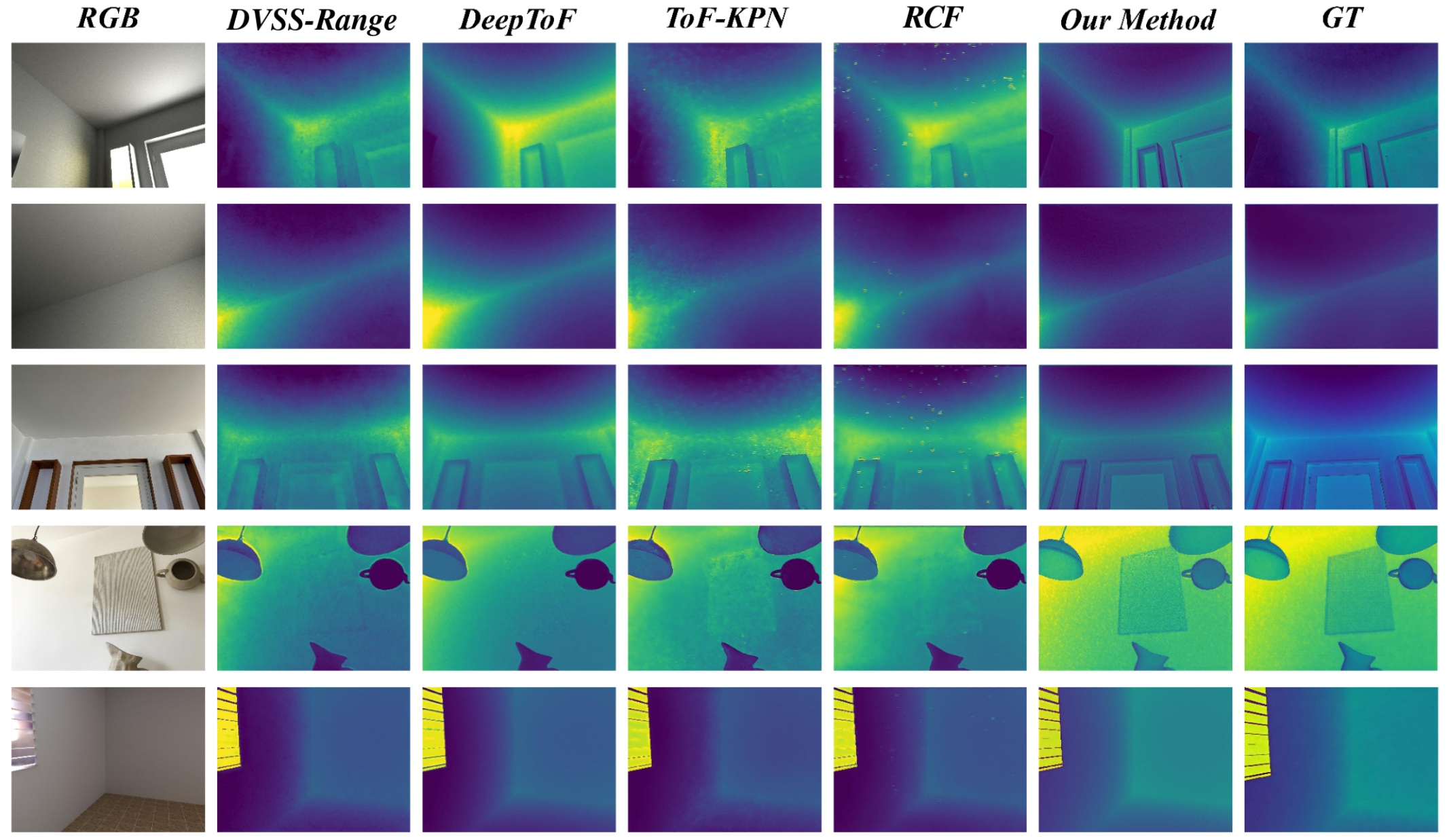}
  \caption{Qualitative comparison on the synthetic test set. From left to right: rendered RGB, DVSS-Range, DeepToF, ToF-KPN, RCF, our method, and ground-truth depth. DVSS-Range denotes the same DVSS backbone trained with range-aware Gaussian noise.}
  \label{fig:fig7}
  \vspace{-10pt}
\end{figure}

Fig.~\ref{fig:fig7} further provides qualitative comparisons on the synthetic test set. DeepToF still exhibits noticeable residual noise and blurred object boundaries. RCF and ToF-KPN reduce part of the noise, but they tend to over-smooth local structures and lose fine geometric details in some regions. In contrast, our method produces cleaner and more spatially consistent depth maps while preserving object contours and structural discontinuities. This behavior is consistent with the design of the DVSS block, where the state-space branch captures long-range contextual dependencies and the local refinement branch enhances fine spatial details. The comparison with the same DVSS backbone trained using range-aware Gaussian noise further shows that the improvement originates not only from the network architecture but also from the proposed training distribution.

\begin{table}[t]
\centering
\captionsetup{font=footnotesize}
\caption{Quantitative comparison with existing iToF denoising methods on the synthetic test set. The proposed method consistently achieves lower reconstruction errors while maintaining higher structural fidelity than the compared baselines.}
\label{tab:quantitative_comparison}
\vspace{1mm}
\setlength{\tabcolsep}{2.6pt}
\renewcommand{\arraystretch}{1.15}
\scriptsize
\begin{tabular}{@{}ccccc@{}}
\toprule
Method
& \makecell[c]{PSNR\\(dB)$\uparrow$}
& \makecell[c]{SSIM\\$\uparrow$}
& \makecell[c]{MSE\\(mm$^2$)$\downarrow$}
& \makecell[c]{MAE\\(mm)$\downarrow$} \\
\midrule
DeepToF
& 37.99
& 0.9951
& 158.94
& 6.12 \\
ToF-KPN
& 39.19
& 0.9969
& 120.52
& 4.88 \\
RCF
& 39.53
& 0.9972
& 111.41
& 4.63 \\
\midrule
Ours w/ $\mathcal{L}_{1}$
& 40.37
& 0.9985
& 91.75
& 2.78 \\
\textbf{Ours w/ $\mathcal{L}_{\mathrm{full}}$}
& \textbf{40.85}
& \textbf{0.9987}
& \textbf{82.30}
& \textbf{2.54} \\
\bottomrule
\end{tabular}
\vspace{-2mm}
\end{table}

\subsection{Ablation Study}

We further study the contribution of the proposed DVSS design and the training objective. As summarized in Table~\ref{tab:ablation_dvss}, the ablation study contains two parts. The first part evaluates the local refinement branch in the DVSS block under the same $\mathcal{L}_{1}$ training objective, while the second part compares different training losses using the full DVSS model.

For the DVSS structure ablation, the ``w/o DWConv \& SCR'' variant removes the entire local refinement branch and only keeps the state-space modeling branch. The ``w/o DWConv'' variant removes the depthwise convolution while retaining SCR. The ``w/ SE'' variant replaces SCR with a standard squeeze-and-excitation module while keeping DWConv. The ``Full DVSS'' variant corresponds to the complete DVSS block equipped with both DWConv and SCR. The results show that removing the whole local refinement branch leads to the largest reconstruction error, suggesting that global dependency modeling alone is insufficient for high-fidelity depth restoration. Adding SCR without DWConv improves PSNR from 39.76~dB to 40.00~dB and reduces MAE from 3.55 to 3.19~mm, indicating that spatial-channel refinement is beneficial. Replacing SCR with SE improves the results over the variant without the local branch, but it remains inferior to the full DVSS block. This comparison suggests that channel-only recalibration is less effective than the proposed spatial-channel refinement for iToF depth denoising.

The full DVSS model trained with $\mathcal{L}_{1}$ achieves the best results among the structure-ablation variants, reaching 40.37~dB PSNR, 0.9985 SSIM, 91.75~mm$^2$ MSE, and 2.78~mm MAE. Compared with the variant without DWConv and SCR, it improves PSNR by 0.61~dB and reduces MSE and MAE by 13.2\% and 21.7\%, respectively. Compared with the ``w/ SE'' variant, it further reduces MSE from 96.35 to 91.75~mm$^2$ and MAE from 3.06 to 2.78~mm. These results validate the complementary roles of DWConv and SCR: DWConv strengthens local spatial interactions, while SCR adaptively refines spatial and channel responses for more accurate depth recovery.

We further evaluate the effect of the training objective on the full DVSS model. The model trained with $\mathcal{L}_{\mathrm{nar}}$ uses the proposed noise-aware reconstruction term, which weights the robust reconstruction penalty according to the propagated depth uncertainty. Compared with the $\mathcal{L}_{1}$-trained full DVSS model, $\mathcal{L}_{\mathrm{nar}}$ improves PSNR from 40.37~dB to 40.66~dB and reduces MSE from 91.75 to 85.90~mm$^2$, indicating that uncertainty-aware supervision helps the network focus on more severely degraded regions. The model trained with $\mathcal{L}_{\mathrm{full}}$ further introduces the gradient consistency term, i.e., $\mathcal{L}_{\mathrm{full}}=\mathcal{L}_{\mathrm{nar}}+\lambda_g\mathcal{L}_{\mathrm{grad}}$. This full objective achieves the best overall performance, reaching 40.85~dB PSNR, 0.9987 SSIM, 82.30~mm$^2$ MSE, and 2.54~mm MAE. Compared with the $\mathcal{L}_{1}$-trained full DVSS model, it improves PSNR by 0.47~dB and reduces MSE and MAE by 10.3\% and 8.6\%, respectively. These results suggest that the uncertainty-aware reconstruction term and the gradient consistency term are complementary: the former emphasizes high-uncertainty low-signal regions, while the latter helps preserve local geometric discontinuities.

\begin{table}[t]
\centering
\captionsetup{font=footnotesize} 
\caption{Ablation study on the proposed DVSS block and training objective. Different structural variants and loss configurations are evaluated to quantify their contributions to depth restoration accuracy and structural preservation.}
\label{tab:ablation_dvss}
\setlength{\tabcolsep}{3.6pt}
\renewcommand{\arraystretch}{1.12}
\scriptsize
\begin{tabular}{lccccc}
\toprule
Variant & Loss & \makecell[c]{PSNR\\(dB)$\uparrow$} & SSIM$\uparrow$ & \makecell[c]{MSE\\(mm$^2$)$\downarrow$} & \makecell[c]{MAE\\(mm)$\downarrow$} \\
\midrule
\multicolumn{6}{c}{\textit{DVSS structure ablation with $\mathcal{L}_{1}$}} \\
\midrule
w/o DWConv \& SCR 
& $\mathcal{L}_{1}$ 
& 39.76 & 0.9954 & 105.66 & 3.55 \\
w/o DWConv 
& $\mathcal{L}_{1}$ 
& 40.00 & 0.9976 & 99.89 & 3.19 \\
w/ SE 
& $\mathcal{L}_{1}$ 
& 40.16 & 0.9983 & 96.35 & 3.06 \\
Full DVSS 
& $\mathcal{L}_{1}$ 
& 40.37 & 0.9985 & 91.75 & 2.78 \\
\midrule
\multicolumn{6}{c}{\textit{Training objective ablation with the full DVSS block}} \\
\midrule
Full DVSS 
& $\mathcal{L}_{\mathrm{nar}}$ 
& 40.66 & 0.9986 & 85.90 & 2.62 \\
Full DVSS 
& $\mathcal{L}_{\mathrm{full}}$ 
& \textbf{40.85} 
& \textbf{0.9987} 
& \textbf{82.30} 
& \textbf{2.54} \\
\bottomrule
\end{tabular}
\vspace{2pt}

{\scriptsize \textit{Note:} ``w/'' and ``w/o'' denote with and without, respectively. ``w/ SE'' means replacing SCR with SE.}
\vspace{-10pt}
\end{table}

\subsection{Real-World Evaluation}

We further conduct real-world depth denoising experiments using the developed iToF prototype to evaluate the practical effectiveness of the proposed method under real imaging conditions. To ensure a fair comparison, all methods are evaluated on observations derived from the same standard-exposure acquisition captured by the prototype. The acquisition configuration, including modulation frequency, exposure setting, and demodulation scheme, is kept fixed across methods, and all scenes are captured under consistent environmental conditions. We include the camera's default phase-shifting demodulation result as a conventional baseline and compare our method with representative learning-based approaches, including DeepToF, ToF-KPN, and RCF.

\begin{figure}[!htbp]
  \centering
  \captionsetup{font=footnotesize}
  \includegraphics[width=0.48\textwidth]{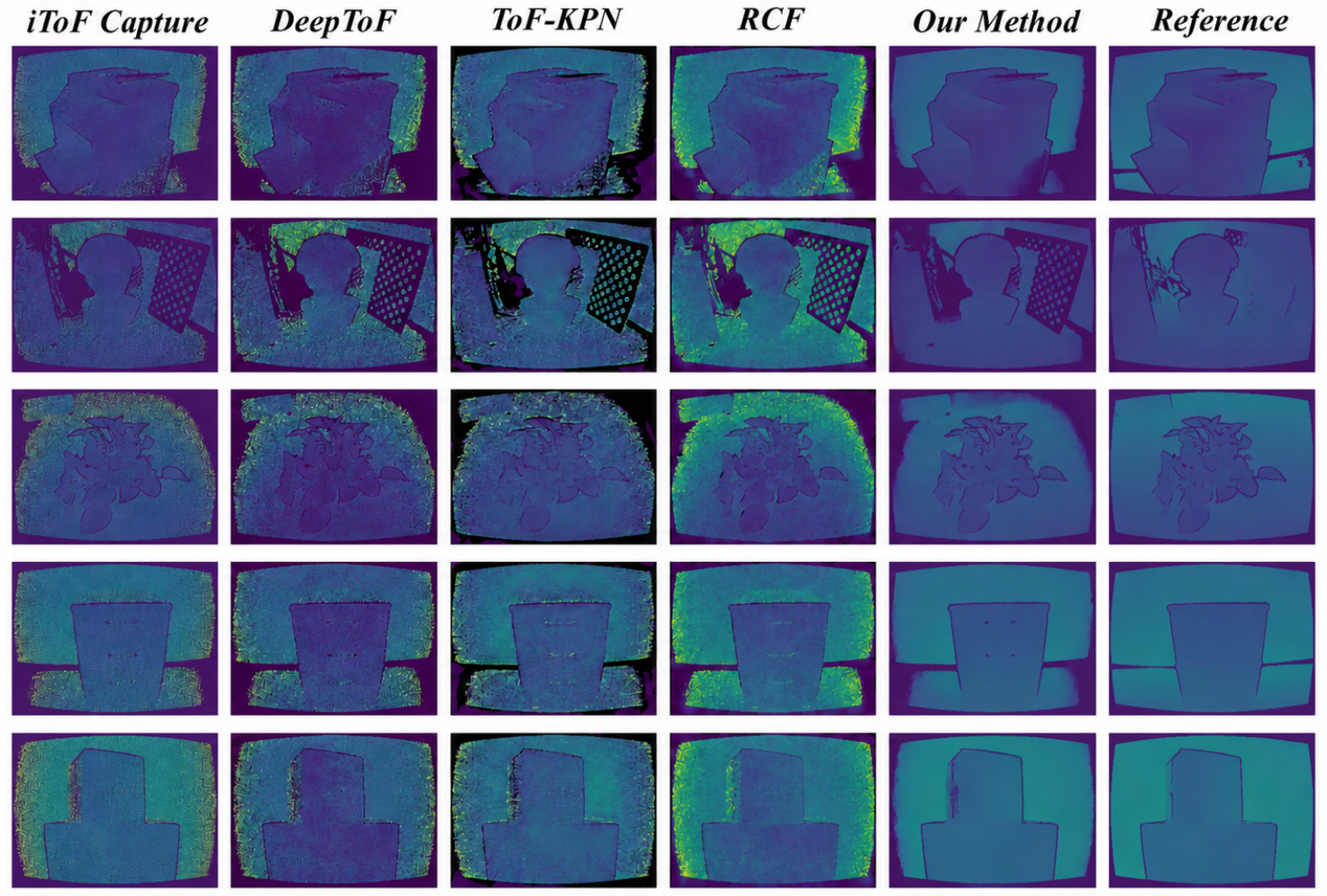}
  \caption{Real-world results of iToF depth denoising. Each row shows one captured scene. From left to right: raw iToF measurement, DeepToF, ToF-KPN, RCF, our method, and the low-noise reference obtained by high-exposure multi-frame averaging. Compared with the baselines, our method more effectively suppresses sensor noise while preserving object boundaries and planar structures.}
  \label{fig:real_world_results}
  \vspace{-10pt}
\end{figure}

For quantitative evaluation on real captured scenes, we construct a low-noise reference depth by high-exposure multi-frame averaging. Specifically, after capturing the standard-exposure four-tap measurements used as the noisy input, we keep the camera pose, scene geometry, illumination condition, modulation frequency, and demodulation scheme unchanged, and capture 100 additional frames under a higher exposure setting while avoiding pixel saturation. These high-exposure frames are processed by the same phase-demodulation and device-level calibration pipeline as the input measurement. The reference depth is obtained by pixel-wise averaging over valid high-exposure depth frames, excluding saturated and invalid pixels. All real-data metrics are computed over the common valid ROI shared by the input, the restored results, and the reference depth. The reference depth is used only for evaluation and is not involved in network training. The real test set contains 200 static scenes. For each scene, one standard-exposure frame is used as the noisy input, while the corresponding low-noise reference is constructed from 100 high-exposure frames. PSNR and MAE are first computed over the common valid ROI of each scene and then averaged across scenes. Fig.~\ref{fig:real_world_results} presents five representative scenes from the test set.

\begin{table}[t]
\centering
\captionsetup{font=footnotesize}
\caption{Quantitative comparison on real iToF measurements. PSNR and MAE are reported to evaluate the restoration accuracy of different methods under practical imaging conditions.}
\label{tab:real_comparison}
\vspace{1mm}
\setlength{\tabcolsep}{8pt}
\renewcommand{\arraystretch}{1.15}
\small
\begin{tabular}{ccc}
\toprule
Method
& PSNR (dB)$\uparrow$
& MAE (mm)$\downarrow$ \\
\midrule
Raw iToF
& 31.68
& 7.21 \\ 

DeepToF
& 33.29
& 6.18 \\ 

ToF-KPN
& 34.76
& 5.31 \\ 

RCF
& 34.48
& 5.52 \\ 

\midrule
\textbf{Ours}
& \textbf{35.42}
& \textbf{4.87} \\
\bottomrule
\end{tabular}
\vspace{-2mm}
\end{table}

As summarized in Table~\ref{tab:real_comparison}, the proposed method achieves the best performance on the real captured data, obtaining 35.42~dB PSNR and 4.87~mm MAE. Compared with the default demodulation baseline, it improves PSNR by 3.74~dB and reduces MAE by 32.5\%. Compared with the strongest learning-based baseline, ToF-KPN, our method further improves PSNR by 0.66~dB and reduces MAE by 8.3\%. Although the absolute performance on real data is lower than that on synthetic data because of more complex sensor noise, non-ideal system responses, and material-dependent reflectance variations, the proposed method maintains the best overall accuracy among all compared approaches. These results demonstrate that the network trained with the proposed heteroscedastic synthesis strategy generalizes well to practical iToF measurements.

The qualitative results in Fig.~\ref{fig:real_world_results} further support the quantitative findings. The default iToF capture contains visible depth fluctuations and structured artifacts. DeepToF reduces part of the noise but leaves residual artifacts around object boundaries. RCF and ToF-KPN produce smoother depth maps, but they may blur local discontinuities and planar structures. In contrast, our method more effectively suppresses depth noise while preserving object boundaries and geometric details. The recovered depth maps show fewer fluctuations on homogeneous surfaces and cleaner transitions near structural edges, leading to more stable and visually coherent reconstructions on real-world scenes.

\subsection{Effect and Generality of the Proposed Noise Model}
\label{sec:noise_model_generality}

To isolate the effect of the training-noise model, we train the same DVSS backbone using fixed Gaussian noise, range-aware Gaussian noise, and the proposed heteroscedastic synthesis. All configurations use the same training samples, network architecture, optimization settings, and standard $\mathcal{L}_{1}$ reconstruction loss. The average perturbation variances of the two Gaussian baselines are matched to that of the proposed model, ensuring that the comparison reflects differences in the spatial noise distribution rather than the overall perturbation magnitude. We further apply the range-aware Gaussian noise and the proposed heteroscedastic synthesis to U-Net~\cite{ronneberger2015unet}, Restormer~\cite{zamir2022restormer}, and DVSS to evaluate whether the benefit of the proposed noise model generalizes across different restoration architectures. For the cross-backbone study, we use a four-stage U-Net with an initial channel width of 32 and a compact Restormer with an initial feature width of 32 and $[2,2,4,4]$ Transformer blocks across four resolution levels. For each backbone, the architecture is kept identical between the range-aware Gaussian and proposed synthesis settings; therefore, the comparison focuses on the within-backbone effect of the training-noise model rather than parameter-matched performance across different architectures.

\begin{table*}[t]
\centering
\captionsetup{font=footnotesize}
\caption{Controlled comparison of training-noise models and their cross-backbone generalization. The results quantify the effectiveness and cross-backbone consistency of the proposed heteroscedastic synthesis under the standard $\mathcal{L}_{1}$ reconstruction loss.}
\label{tab:noise_model_generality}
\setlength{\tabcolsep}{7pt}
\renewcommand{\arraystretch}{1.12}
\small
\begin{tabular}{cccccc}
\toprule
Backbone
& Training-noise model
& \makecell[c]{Synthetic PSNR\\(dB)$\uparrow$}
& \makecell[c]{Synthetic MAE\\(mm)$\downarrow$}
& \makecell[c]{Real PSNR\\(dB)$\uparrow$}
& \makecell[c]{Real MAE\\(mm)$\downarrow$} \\
\midrule
\multicolumn{6}{c}{\textit{Noise-model ablation on the DVSS backbone}} \\
\midrule
DVSS
& Fixed Gaussian noise
& 39.40
& 3.42
& 34.41
& 5.82 \\

DVSS
& Range-aware Gaussian noise
& 39.89
& 3.08
& 34.88
& 5.36 \\

\textbf{DVSS}
& \textbf{Proposed heteroscedastic synthesis}
& \textbf{40.37}
& \textbf{2.78}
& \textbf{35.12}
& \textbf{5.04} \\

\midrule
\multicolumn{6}{c}{\textit{Cross-backbone generalization of the proposed synthesis}} \\
\midrule
U-Net
& Range-aware Gaussian noise
& 38.56
& 4.18
& 33.67
& 6.34 \\

U-Net
& Proposed heteroscedastic synthesis
& 39.01
& 3.76
& 34.12
& 5.84 \\

\addlinespace[1pt]

Restormer
& Range-aware Gaussian noise
& 39.22
& 3.58
& 34.19
& 5.91 \\

Restormer
& Proposed heteroscedastic synthesis
& 39.68
& 3.22
& 34.55
& 5.46 \\

\addlinespace[1pt]

DVSS
& Range-aware Gaussian noise
& 39.89
& 3.08
& 34.88
& 5.36 \\

\textbf{DVSS}
& \textbf{Proposed heteroscedastic synthesis}
& \textbf{40.37}
& \textbf{2.78}
& \textbf{35.12}
& \textbf{5.04} \\
\bottomrule
\end{tabular}
\vspace{-3mm}
\end{table*}

As shown in Table~\ref{tab:noise_model_generality}, training with fixed Gaussian noise yields 39.40~dB PSNR and 3.42~mm MAE on the synthetic test set, together with 34.41~dB PSNR and 5.82~mm MAE on real measurements. Introducing distance-dependent noise improves the corresponding results to 39.89~dB, 3.08~mm, 34.88~dB, and 5.36~mm. The proposed heteroscedastic synthesis further improves the synthetic performance to 40.37~dB PSNR and 2.78~mm MAE, while obtaining 35.12~dB PSNR and 5.04~mm MAE on real measurements. These results show that range dependence provides a more realistic training distribution than spatially uniform Gaussian noise, while additionally incorporating returned-signal strength and tap-response sensitivity further reduces the discrepancy between synthetic and real measurements.

The cross-backbone results exhibit a consistent trend. Replacing range-aware Gaussian noise with the proposed heteroscedastic synthesis reduces the synthetic MAE from 4.18 to 3.76~mm for U-Net, from 3.58 to 3.22~mm for Restormer, and from 3.08 to 2.78~mm for DVSS. On real measurements, the corresponding MAEs decrease from 6.34 to 5.84~mm, from 5.91 to 5.46~mm, and from 5.36 to 5.04~mm, respectively. The consistent improvements across convolutional, Transformer, and state-space restorers demonstrate that the proposed synthesis is not tailored to a particular network architecture, but instead provides a sensor-matched training distribution that benefits different restoration backbones.

Among the evaluated backbones, DVSS achieves the best absolute performance under both training-noise settings, indicating that the proposed heteroscedastic synthesis and the global-local DVSS architecture provide complementary improvements. Furthermore, replacing the standard $\mathcal{L}_{1}$ objective with the proposed uncertainty-aware reconstruction and gradient-consistency terms improves the DVSS model from 40.37 to 40.85~dB on the synthetic test set and from 35.12 to 35.42~dB on real measurements, while reducing the corresponding MAEs from 2.78 to 2.54~mm and from 5.04 to 4.87~mm.

\section{Conclusion and Future Work}

This paper presents a sensor-uncertainty-aware framework for reliable iToF depth sensing by coupling tap-level noise modeling, heteroscedastic noisy-depth synthesis, and learning-based depth restoration. Starting from depth-dependent tap responses and returned-signal-dependent measurement statistics, we derive a pixel-wise depth uncertainty model through a depth-oriented weighted least-squares formulation. The resulting uncertainty is used for both heteroscedastic training-data synthesis and uncertainty-aware restoration supervision. Based on this data construction strategy, the proposed DVSS network combines state-space global modeling with convolutional spatial-channel refinement to recover accurate and structure-preserving depth. Experiments on synthetic data and real measurements from an in-house iToF prototype validate the proposed uncertainty model under varying range and returned-signal conditions. Controlled comparisons with fixed and range-aware Gaussian noise, together with cross-backbone evaluations on U-Net, Restormer, and DVSS, further demonstrate that the proposed synthesis consistently improves different restoration architectures. The complete framework achieves the best synthetic and real depth accuracy among the evaluated methods. Future work will extend the framework to systematic non-ideal factors such as multipath interference, motion artifacts, and thermal drift, and investigate its deployment efficiency on embedded sensing platforms.




\bibliographystyle{IEEEtran}
\bibliography{ref}

@article{kazerouni2022survey,
  title={A survey of state-of-the-art on visual SLAM},
  author={Kazerouni, Iman Abaspur and Fitzgerald, Luke and Dooly, Gerard and Toal, Daniel},
  journal={Expert Systems with Applications},
  volume={205},
  pages={117734},
  year={2022},
  publisher={Elsevier}
}

@article{du2025modeling,
  title={Modeling, analysis, and optimization of random error in indirect time-of-flight camera},
  author={Du, Yansong and Jiang, Zhaoxiang and Tian, Jindong and Guan, Xun},
  journal={Optics Express},
  volume={33},
  number={2},
  pages={1983--1994},
  year={2025},
  publisher={Optica Publishing Group}
}

@article{du2026structure,
  author  = {Du, Y. and Deng, Y. and Zhou, Y. and others},
  title   = {A Structure-Aware Compressed Sensing Framework for Robust Multipath Suppression in Indirect ToF Imaging},
  journal = {IEEE Transactions on Instrumentation and Measurement},
  year    = {2026}
}

@inproceedings{du2025new,
  title={A New Method for Removing Internal Scattering Noise in iToF Camera},
  author={Du, Yansong and Yao, Jingtong and Jiao, Feiyu and Zhou, Yuting and Jin, Qiang and Wang, Bangyao and An, Kang and Jiang, Zhaoxiang and Guan, Xun},
  booktitle={2025 Conference on Lasers and Electro-Optics Europe \& European Quantum Electronics Conference (CLEO/Europe-EQEC)},
  pages={1--1},
  year={2025},
  organization={IEEE}
}

@article{foix2011lockin,
  title     = {Lock-in Time-of-Flight ({ToF}) Cameras: A Survey},
  author    = {Foix, Sergi and Aleny{\`a}, Guillem and Torras, Carme},
  journal   = {IEEE Sensors Journal},
  volume    = {11},
  number    = {9},
  pages     = {1917--1926},
  year      = {2011},
  doi       = {10.1109/JSEN.2010.2101060},
  publisher = {IEEE}
}

@article{illade2015shotnoise,
  title     = {Distance Measurement Error in Time-of-Flight Sensors Due to Shot Noise},
  author    = {Illade-Quinteiro, Javier and Brea, V{\'i}ctor M. and L{\'o}pez, Paula and Cabello, David and Dom{\'e}nech-Asensi, Gin{\'e}s},
  journal   = {Sensors},
  volume    = {15},
  number    = {3},
  pages     = {4624--4642},
  year      = {2015},
  doi       = {10.3390/s150304624},
  publisher = {MDPI}
}

@article{lambers2015simulation,
  title     = {Simulation of Time-of-Flight Sensors for Evaluation of Chip Layout Variants},
  author    = {Lambers, Martin and Hoberg, Stefan and Kolb, Andreas},
  journal   = {IEEE Sensors Journal},
  volume    = {15},
  number    = {7},
  pages     = {4019--4026},
  year      = {2015},
  doi       = {10.1109/JSEN.2015.2409816},
  publisher = {IEEE}
}

@article{bulczak2018quantified,
  title     = {Quantified, Interactive Simulation of {AMCW} {ToF} Camera Including Multipath Effects},
  author    = {Bulczak, Dominik and Lambers, Martin and Kolb, Andreas},
  journal   = {Sensors},
  volume    = {18},
  number    = {1},
  pages     = {13},
  year      = {2018},
  doi       = {10.3390/s18010013},
  publisher = {MDPI}
}

@article{gao2023deblurring,
  title     = {A Deblurring Method for Indirect Time-of-Flight Depth Sensor},
  author    = {Gao, Jing and Gao, Xueqiang and Nie, Kaiming and Gao, Zhiyuan and Xu, Jiangtao},
  journal   = {IEEE Sensors Journal},
  volume    = {23},
  number    = {3},
  pages     = {2718--2726},
  year      = {2023},
  doi       = {10.1109/JSEN.2022.3229687},
  publisher = {IEEE}
}

@article{wang2022attentiongan,
  title     = {Attention {GAN} for Multipath Error Removal From {ToF} Sensors},
  author    = {Wang, Xin and Zhou, Wenbiao and Jia, Yunfei},
  journal   = {IEEE Sensors Journal},
  volume    = {22},
  number    = {20},
  pages     = {19713--19721},
  year      = {2022},
  doi       = {10.1109/JSEN.2022.3203759},
  publisher = {IEEE}
}

@inproceedings{gutierrez2019practical,
  title     = {Practical Coding Function Design for Time-of-Flight Imaging},
  author    = {Gutierrez-Barragan, Felipe and Reza, Syed Azer and Velten, Andreas and Gupta, Mohit},
  booktitle = {Proceedings of the IEEE/CVF Conference on Computer Vision and Pattern Recognition ({CVPR})},
  pages     = {1566--1574},
  year      = {2019},
  doi       = {10.1109/CVPR.2019.00166}
}

@inproceedings{li2022fisher,
  title     = {Fisher Information Guidance for Learned Time-of-Flight Imaging},
  author    = {Li, Jiaqu and Yue, Tao and Zhao, Sijie and Hu, Xuemei},
  booktitle = {Proceedings of the IEEE/CVF Conference on Computer Vision and Pattern Recognition ({CVPR})},
  pages     = {16334--16343},
  year      = {2022},
  doi       = {10.1109/CVPR52688.2022.01585}
}

@article{marco2017deeptof,
  title     = {{DeepToF}: Off-the-Shelf Real-Time Correction of Multipath Interference in Time-of-Flight Imaging},
  author    = {Marco, Julio and Hernandez, Quercus and Mu{\~n}oz, Adolfo and Dong, Yue and Jarabo, Adrian and Kim, Min H. and Tong, Xin and Gutierrez, Diego},
  journal   = {ACM Transactions on Graphics},
  volume    = {36},
  number    = {6},
  pages     = {219:1--219:12},
  year      = {2017},
  doi       = {10.1145/3130800.3130884},
  publisher = {ACM}
}

@inproceedings{su2018deeptof,
  title     = {Deep End-to-End Time-of-Flight Imaging},
  author    = {Su, Shuochen and Heide, Felix and Wetzstein, Gordon and Heidrich, Wolfgang},
  booktitle = {Proceedings of the IEEE Conference on Computer Vision and Pattern Recognition ({CVPR})},
  pages     = {6383--6392},
  year      = {2018},
  doi       = {10.1109/CVPR.2018.00668}
}

@inproceedings{guo2018flat,
  title     = {Tackling {3D} {ToF} Artifacts Through Learning and the {FLAT} Dataset},
  author    = {Guo, Qi and Frosio, Iuri and Gallo, Orazio and Zickler, Todd and Kautz, Jan},
  booktitle = {Proceedings of the European Conference on Computer Vision ({ECCV})},
  pages     = {368--383},
  year      = {2018},
  doi       = {10.1007/978-3-030-01246-5_23}
}

@inproceedings{agresti2019uda,
  title     = {Unsupervised Domain Adaptation for {ToF} Data Denoising with Adversarial Learning},
  author    = {Agresti, Gianluca and Schaefer, Henrik and Sartor, Piergiorgio and Zanuttigh, Pietro},
  booktitle = {Proceedings of the IEEE/CVF Conference on Computer Vision and Pattern Recognition ({CVPR})},
  pages     = {5584--5593},
  year      = {2019},
  doi       = {10.1109/CVPR.2019.00573}
}

@article{gutierrez2021itof2dtof,
  title     = {{iToF2dToF}: A Robust and Flexible Representation for Data-Driven Time-of-Flight Imaging},
  author    = {Gutierrez-Barragan, Felipe and Chen, Huaijin and Gupta, Mohit and Velten, Andreas and Gu, Jinwei},
  journal   = {IEEE Transactions on Computational Imaging},
  volume    = {7},
  pages     = {1205--1214},
  year      = {2021},
  doi       = {10.1109/TCI.2021.3126533},
  publisher = {IEEE}
}

@inproceedings{schelling2022radu,
  title     = {{RADU}: Ray-Aligned Depth Update Convolutions for {ToF} Data Denoising},
  author    = {Schelling, Michael and Hermosilla, Pedro and Ropinski, Timo},
  booktitle = {Proceedings of the IEEE/CVF Conference on Computer Vision and Pattern Recognition ({CVPR})},
  pages     = {661--670},
  year      = {2022}
}

@article{zhang2024dcs2noise,
  title     = {Non-Systematic Noise Reduction Framework for {ToF} Camera},
  author    = {Zhang, Wuyang and Song, Ping and Bai, Yunjian and Geng, Haocheng and Wu, Yinpeng and Zheng, Zhaolin},
  journal   = {Optics and Lasers in Engineering},
  volume    = {180},
  pages     = {108324},
  year      = {2024},
  doi       = {10.1016/j.optlaseng.2024.108324},
  publisher = {Elsevier}
}

@incollection{dong2024dualcorrelation,
  title     = {Exploiting Dual-Correlation for Multi-Frame Time-of-Flight Denoising},
  author    = {Dong, Guanting and Zhang, Yueyi and Sun, Xiaoyan and Xiong, Zhiwei},
  booktitle = {Computer Vision -- {ECCV} 2024},
  pages     = {473--489},
  year      = {2024},
  publisher = {Springer},
  doi       = {10.1007/978-3-031-72670-5_27}
}

@inproceedings{wang2025giga,
  title     = {Consistent Time-of-Flight Depth Denoising via Graph-Informed Geometric Attention},
  author    = {Wang, Weida and He, Changyong and Zeng, Jin and Qiu, Di},
  booktitle = {Proceedings of the IEEE/CVF International Conference on Computer Vision ({ICCV})},
  year      = {2025},
  note      = {Also available as arXiv:2506.23542}
}

@inproceedings{an2025mpi,
  title        = {{MPI-Mamba}: Cross Propagation Mamba for Multipath Interference Correction},
  author       = {An, Kang and Jiang, Zhaoxiang and Tian, Jindong},
  booktitle    = {2025 IEEE International Conference on Robotics and Automation ({ICRA})},
  pages        = {12774--12781},
  year         = {2025},
  doi          = {10.1109/ICRA55743.2025.11127297},
  organization = {IEEE}
}

@inproceedings{tang2022ghostnetv2,
  title     = {{GhostNetV2}: Enhance Cheap Operation with Long-Range Attention},
  author    = {Tang, Yehui and Han, Kai and Guo, Jianyuan and Xu, Chang and Xu, Chao and Wang, Yunhe},
  booktitle = {Advances in Neural Information Processing Systems},
  volume    = {35},
  pages     = {9969--9982},
  year      = {2022}
}

@inproceedings{cai2023efficientvit,
  title     = {{EfficientViT}: Lightweight Multi-Scale Attention for High-Resolution Dense Prediction},
  author    = {Cai, Han and Li, Junyan and Hu, Muyan and Gan, Chuang and Han, Song},
  booktitle = {Proceedings of the IEEE/CVF International Conference on Computer Vision ({ICCV})},
  pages     = {17302--17313},
  year      = {2023},
  doi       = {10.1109/ICCV51070.2023.01587}
}

@misc{gu2024mamba,
  title         = {Mamba: Linear-Time Sequence Modeling with Selective State Spaces},
  author        = {Gu, Albert and Dao, Tri},
  year          = {2023},
  eprint        = {2312.00752},
  archivePrefix = {arXiv},
  primaryClass  = {cs.LG}
}

@article{sousa2023systematic,
  title     = {A Systematic Literature Review on Long-Term Localization and Mapping for Mobile Robots},
  author    = {Sousa, Ricardo B. and Sobreira, H{\'e}ber M. and Moreira, A. Paulo},
  journal   = {Journal of Field Robotics},
  volume    = {40},
  number    = {5},
  pages     = {1245--1322},
  year      = {2023},
  doi       = {10.1002/rob.22170},
  publisher = {Wiley}
}

@article{dong2024recurrent,
  title={Recurrent cross-modality fusion for time-of-flight depth denoising},
  author={Dong, Guanting and Zhang, Yueyi and Sun, Xiaoyan and Xiong, Zhiwei},
  journal={IEEE Transactions on Computational Imaging},
  volume={10},
  pages={1626--1639},
  year={2024},
  publisher={IEEE}
}

@article{qiao2024rgb,
  title     = {{RGB} Guided {ToF} Imaging System: A Survey of Deep Learning-Based Methods},
  author    = {Qiao, Xin and Poggi, Matteo and Deng, Pengchao and Wei, Hao and Ge, Chenyang and Mattoccia, Stefano},
  journal   = {International Journal of Computer Vision},
  volume    = {132},
  number    = {11},
  pages     = {4954--4991},
  year      = {2024},
  doi       = {10.1007/s11263-024-02089-5},
  publisher = {Springer}
}

@inproceedings{du2025random,
  title     = {Random Phase Noise Optimization for {iToF} Camera},
  author    = {Du, Yansong and Jiang, Zhaoxiang and Yao, Jingtong and Jiao, Feiyu and Wang, Bangyao and Xu, Zhancong and Jin, Qiang and Ma, Yingjia and Guan, Xun},
  booktitle = {{CLEO}: Applications and Technology},
  pages     = {AA118\_7},
  year      = {2025}
}

@article{du2025calibration,
  title   = {A Calibration Method for Indirect Time-of-Flight Cameras to Eliminate Internal Scattering Interference},
  author  = {Du, Yansong and Yao, Jingtong and Zhou, Yuting and Jiao, Feiyu and Jiang, Zhaoxiang and Guan, Xun},
  journal = {arXiv preprint arXiv:2511.01874},
  year    = {2025}
}

@inproceedings{deng2025monocular,
  title        = {Monocular Depth Estimation Assisted {iToF}-{RGB} Fusion for Improved Depth Resolution},
  author       = {Deng, Yutong and Du, Yansong and Zhou, Yuting and Jiao, Feiyu and Jian, Song and Guan, Xun},
  booktitle    = {2025 Asia Communications and Photonics Conference ({ACP})},
  pages        = {1--5},
  year         = {2025},
  organization = {IEEE}
}

@inproceedings{song2022all,
  title     = {All the Attention You Need: Global-Local, Spatial-Channel Attention for Image Retrieval},
  author    = {Song, Chun-Ho and Han, Hyeong-Joon and Avrithis, Yannis},
  booktitle = {Proceedings of the IEEE/CVF Winter Conference on Applications of Computer Vision ({WACV})},
  pages     = {2754--2763},
  year      = {2022}
}

@article{wang2025mamba,
  title     = {Is Mamba Effective for Time Series Forecasting?},
  author    = {Wang, Zihan and Kong, Fuxi and Feng, Shi and Wang, Ming and Yang, Xiaocui and Zhao, Hao and Wang, Daling and Zhang, Yifei},
  journal   = {Neurocomputing},
  volume    = {619},
  pages     = {129178},
  year      = {2025},
  publisher = {Elsevier}
}

@inproceedings{qiu2019deep,
  title     = {Deep End-to-End Alignment and Refinement for Time-of-Flight {RGB-D} Module},
  author    = {Qiu, Dongxiao and Pang, Jiangmiao and Sun, Wenxiu and Yang, Chengxi},
  booktitle = {Proceedings of the IEEE/CVF International Conference on Computer Vision ({ICCV})},
  pages     = {9994--10003},
  year      = {2019}
}

@article{idoughi2026bayestof,
  title={{BayesToF}: Multiresolution Denoising of Indirect Time-of-Flight Distance Maps},
  author={Idoughi, Achour and Raffoul, Joseph and Hirakawa, Keigo},
  journal={Optics Express},
  volume={34},
  number={11},
  pages={20044--20065},
  year={2026},
  doi={10.1364/OE.596035},
  publisher={Optica Publishing Group}
}

@inproceedings{ronneberger2015unet,
  title={U-Net: Convolutional Networks for Biomedical Image Segmentation},
  author={Ronneberger, Olaf and Fischer, Philipp and Brox, Thomas},
  booktitle={Medical Image Computing and Computer-Assisted Intervention (MICCAI)},
  pages={234--241},
  year={2015},
  publisher={Springer}
}

@inproceedings{zamir2022restormer,
  title={Restormer: Efficient Transformer for High-Resolution Image Restoration},
  author={Zamir, Syed Waqas and Arora, Aditya and Khan, Salman and Hayat, Munawar and Khan, Fahad Shahbaz and Yang, Ming-Hsuan},
  booktitle={Proceedings of the IEEE/CVF Conference on Computer Vision and Pattern Recognition},
  pages={5728--5739},
  year={2022}
}
\end{document}